\documentclass[aps,prl,twocolumn,showpacs,floatfix]{revtex4}
\usepackage[latin1]{inputenc}
\usepackage{bbm,slashed}
\usepackage{graphicx,epsfig}
\usepackage{amsmath,amsfonts,amssymb}

\newcommand{\N}{\mathcal{N}}  
\newcommand{\al}{\alpha}
\newcommand{\Zb}{\mathbb Z}

\newcommand{\Eqref}[1]{Eq.~\eqref{#1}}

\DeclareMathOperator{\STr}{STr}

\DeclareMathAlphabet{\boldmathe}{T1}{cmr}{bx}{it}
\newcommand{\abs}[1]{\left| #1 \right|}

\newcommand{\psibar}{\overline{\psi}}

\begin{document}
\title{Supersymmetry breaking as a quantum phase transition} 
\author{Holger Gies, Franziska Synatschke and Andreas Wipf}
\affiliation{
Theoretisch-Physikalisches Institut, Friedrich-Schiller-Universit{\"a}t
Jena,
Max-Wien-Platz 1, D-07743 Jena, Germany}

\begin{abstract}
We explore supersymmetry breaking in the light of a rich fixed-point
structure of two-dimensional supersymmetric Wess-Zumino models with one
supercharge using the functional renormalization group (RG). We relate the
dynamical breaking of supersymmetry to an RG relevant control parameter of the
superpotential which is a common relevant direction of all fixed points of the
system. Supersymmetry breaking can thus be understood as a quantum phase
transition analogous to similar transitions in correlated fermion
systems. Supersymmetry gives rise to a new superscaling relation between the
critical exponent associated with the control parameter and the anomalous
dimension of the field -- a scaling relation which is not known in standard
spin systems.  
\end{abstract}
\maketitle

Supersymmetry is an important guiding principle for the construction of models
beyond the standard model of particle physics, as it helps parametrizing the
problem of a large hierarchy of scales, supports the unification of gauge
couplings, and facilitates an intricate combination of internal symmetries
with the Poincar\'{e} group. On the other hand, supersymmetry must be broken
in nature in a specific way without violating these attractive features and
other constraints imposed by the standard-model precision data. In addition to
a variety of breaking mechanisms often involving further hidden gauge and
particle sectors, dynamical supersymmmetry breaking \cite{Witten:1981nf} can
play a decisive role in the formation of the low-energy standard
model. Generically, symmetry breaking is related to collective condensation
phenomena, often requiring a nonperturbative understanding of the
fluctuation-induced dynamics.

From the viewpoint of statistical physics, symmetry breaking and phase
transitions between ground states of a differently realized symmetry are often
related to fixed points of the renormalization group (RG). In addition, the
quantitative properties of the system near a phase transition are determined
by the RG flow in the fixed-point regime. As the understanding of symmetry
breaking phenomena in quantum field theory and particle physics has often
profited from analogies to statistical systems, this paper
explores supersymmetry breaking in the light of phase transitions and
critical phenomena.

As a simple example, we concentrate on Wess-Zumino models with $\N=1$
supercharges in two dimensions in Euclidean spacetime. Dynamical supersymmetry
breaking in this and many other supersymmetric systems is governed by a
control parameter of the superpotential which plays the role of a bosonic
mass term. This is very reminiscent to phase transitions in strongly
correlated fermionic systems: here,  symmetry breaking is governed by a
composite bosonic order parameter, e.g., a superfluid condensate, the mass
term of which serves as the control parameter of the phase transition. For
instance, the inverse interaction strength of  attractively interacting
electrons on a honeycomb lattice governs the phase transition
between the semimetallic and the superfluid phase as discussed in the context
of ultracold atoms \cite{zhao:230404} or graphene \cite{uchoa:146801}. As this
transition occurs at zero temperature, the critical value of the control
parameter marks a quantum critical point of a quantum phase transition. 

A standard approach to quantum phase transitions in strongly correlated
fermion systems is the Hertz-Millis theory
\cite{PhysRevB.14.1165,PhysRevB.48.7183} where composite bosonic fields are
introduced by a Hubbard-Stratonovich transformation. Subsequently, the
fermionic degrees of freedom are integrated out, leaving a purely bosonic
description which is dealt with in a local approximation. As this strategy is
not always sufficient, a treatment of fundamental fermionic and composite
bosonic degrees of freedom on equal footing beyond the Hertz-Millis theory has
recently proved successful \cite{strack-2009}.

Now, supersymmetric systems by definition are combined bosonic-fermionic
systems with a high degree of symmetry. A treatment of these degrees of
freedom on equal footing is even mandatory to preserve supersymmetry
manifestly. Apart from this symmetry, we will show that
supersymmetry breaking has many similarities to a quantum phase transition in
strongly correlated fermion systems. Beyond these similarities, supersymmetry
invokes additional structures which go beyond those known in statistical
physics.

Euclidean two-dimensional $\N=1$ Wess-Zumino models can be defined by the
Lagrange density in the off-shell formulation $\mathcal L=\mathcal
L_\text{kin}+\mathcal L_\text{int}$ with
\begin{align}
	\mathcal L_\text{kin}=&
	\frac12\partial_\mu\phi\partial^\mu\phi+
	\frac i4\psibar\slashed{\partial}\psi-
	\frac i4\partial_\mu\psibar\gamma^\mu\psi
	-\frac12F^2 ,\label{eq:kin}\\
	\mathcal
        L_\text{int}=&\frac12W''(\phi)\psibar\gamma_\ast\psi-W'(\phi)F \label{eq:int},
\end{align}
with superpotential $W(\phi)$, bosonic field $\phi$, fermionic fields $\psi$
and $\psibar$ and auxiliary field $F$. The prime denotes the derivative with
respect to $\phi$, and we use a chiral representation for the $\gamma$
matrices. This model allows for dynamical supersymmetry breaking if the
highest power of the bare superpotential is odd.

We use the functional RG to calculate the effective action $\Gamma$ of the
theory. The functional RG can be formulated as a flow equation for the {\em
  effective average action} $\Gamma_k$ \cite{Wetterich:1992yh}. This is a
scale-dependent action functional which interpolates between the classical
action $S=\int(\mathcal{L}_{\text{kin}}+\mathcal{L}_{\text{int}})$ and the
full quantum effective action $\Gamma$. The interpolation scale $k$ denotes an
infrared (IR) regulator scale, such that no fluctuations are included for
$k\to\Lambda$ (with $\Lambda$ being the UV cutoff), implying
$\Gamma_{k\to \Lambda}\to S$. For $k\to0$, all fluctuations are taken into
account such that $\Gamma_{k\to0}\to \Gamma$ is the full quantum solution. The
effective average action can be determined from the Wetterich equation
\cite{Wetterich:1992yh}
\begin{equation}
 \partial_t\Gamma_k=
 \frac12 \STr\left\{\left[\Gamma_k^{(2)}+ R_k\right]^{-1}\partial_t
 R_k\right\}, \quad t=\ln\frac{k}{\Lambda},
\label{eq:wetterich}
\end{equation}
which defines an RG flow trajectory in theory space with $S$ serving as
initial condition. Here, $
\left(\Gamma_k^{(2)}\right)_{ab}=\frac{\overrightarrow{\delta}}{\delta\Psi_a}
\Gamma_k\frac {\overleftarrow{\delta}}{\delta\Psi_b}$ where the indices $a,b$
summarize field components, internal and Lorentz indices, as well as spacetime
or momentum coordinates, i.e., $\Psi^{\text T}=(\phi,F,\psi,\bar\psi)$. The
regulator function $R_k$ is introduced in the form of a quadratic contribution
to the classical action, $\Delta S_k=\int \Psi^{\text{T}} R_k \Psi$. $R_k$ can
thus be viewed as a momentum-dependent mass term, implementing the IR
suppression of modes below $k$. Different functional forms of $R_k$ correspond
to different RG schemes. Physical quantities do not depend on the regulator.
  
For a manifestly supersymmetric flow equation, we derive the regulator from a
$D$ term of a quadratic superfield operator, similarly to our construction for
supersymmetric quantum mechanics \cite{Synatschke:2008pv}.  The most general
supersymmetric cutoff action in the off-shell formulation in momentum space is
given by 
\begin{equation*}
 \Delta S_k\!=\!\frac12\!\int \!d^2{p} (r_1\psibar\gamma_\ast\psi -2r_1\phi
F+p^2 r_2\phi^2+r_2\psibar\slashed p\psi-r_2F^2).
\end{equation*}
The
regulator shape functions $r_{1,2}=r_{1,2}(p^2)$ determine the
precise form of the Wilsonian momentum shell integration.  The form of $\Delta
S_k$ guarantees that the whole RG trajectory remains in the hypersurface of
supersymmetric action functionals even within approximative calculations; see
\cite{Sonoda:2008dz,Rosten:2008ih} for further supersymmetric RG studies.

In general, the full quantum effective action consists of all possible
operators which are compatible with the symmetries of the theory. In this
paper, we use a supercovariant derivative expansion to systematically
classify these operators. Truncating the expansion at a given order
constitutes a consistent approximation scheme. In a variety of bosonic and
fermionic systems, such expansions have been proven to capture the physics of phase
transition and critical phenomena quantitatively, see,
e.g. \cite{Delamotte-2007}. For our quantitative results presented
  below, we observe a satisfactory convergence: leading-order results
  generically receive $\mathcal O(1-10\%)$ corrections at next-to-leading
  order, increasing up to $\sim 50\%$ for observables in the infinite-coupling limit.

Let us start with the lowest-order derivative expansion, corresponding to a
local-potential approximation for the superpotential. In this truncation, the
effective action is approximated by 
\begin{equation}
\Gamma_k=\int d\tau\left[\mathcal
  L_\text{kin} +\frac12W''_k(\phi)\psibar\gamma_\ast\psi-W'_k(\phi)F\right]
\end{equation}
with a $k$-dependent superpotential $W_k$. Projecting the flow equation
\eqref{eq:wetterich} onto this truncation leads to the flow equation for the
superpotential. For a simplified regulator choice with  $r_1=0$
and $r_2=\left(\abs{{k}/{p}}-1\right)\theta\left(1-({p^2}/{k^2})\right)$, the
superpotential flow reduces to
\begin{align} 
	\partial_tW_k(\phi)
	=&-\frac{k^2}{4\pi}
	\frac{W''_k(\phi)}{k^2+W''_k(\phi)^2}.
	\label{eq:flow3}
\end{align}
This regulator choice is convenient for analytical studies. We have verified
explicitly by using a variety of different regulators that all conclusions
drawn below from this simple choice also hold for other regulators. 

The corresponding flow equation for the dimensionless potential $w_k=W_k/k$
admits a variety of nontrivial fixed-point solutions: already a polynomial
expansion for $\Zb_2$ antisymmetric superpotentials to order $n$, 
\begin{equation}
w_k'\equiv \frac{\partial w_k}{\partial\phi}=\lambda (\phi^2-a^2) + b_4
\phi^4+b_6\phi^6+\dots+b_{2n}\phi^{2n},\label{eq:poly}
\end{equation}
reveal $n$ independent solutions to the fixed-point equation $\partial_t
w_k=0$, in addition to the Gau\ss ian fixed point\footnote{Beyond the
  polynomial expansion, we find fixed-point solutions corresponding to
  oscillating sine-Gordon-type potentials, bouncing potentials as well as
  potentials which confine the field to a compact target space; see
  \cite{Synatschke:2009} for a detailed discussion.}. The latter has
  infinitely many relevant directions in agreement with perturbative power-counting in two
dimensions. The number of relevant RG directions decreases for the other fixed
points, culminating in a maximally IR-stable fixed point with only one
relevant direction. This relevant direction exactly corresponds to the
parameter $a^2$ in the superpotential, and it is common to all other fixed
points as well.  As $a^2$ is related to the bosonic mass term, we introduce a
critical exponent $\nu_W$ for this direction, such that $a^2$ scales as
$a^2\sim k^{-1/\nu_W}$ at a fixed point. This critical exponent thus
corresponds to the negative inverse of the associated eigenvalue of the
stability matrix. The exponent $\nu_W$ plays a similar role for the
superpotential, as the exponent $\nu$ for the effective potential in
Ising-type systems. At leading order in the derivative expansion, we obtain
from \Eqref{eq:flow3}: $\nu_W|_{\text{LO}}=1$.  Fixed-point potentials and
critical exponents at next-to-leading order will be discussed below.

\begin{figure} 
\vspace*{-2.5\baselineskip}
\centering{ 
\includegraphics[width=.99\columnwidth]{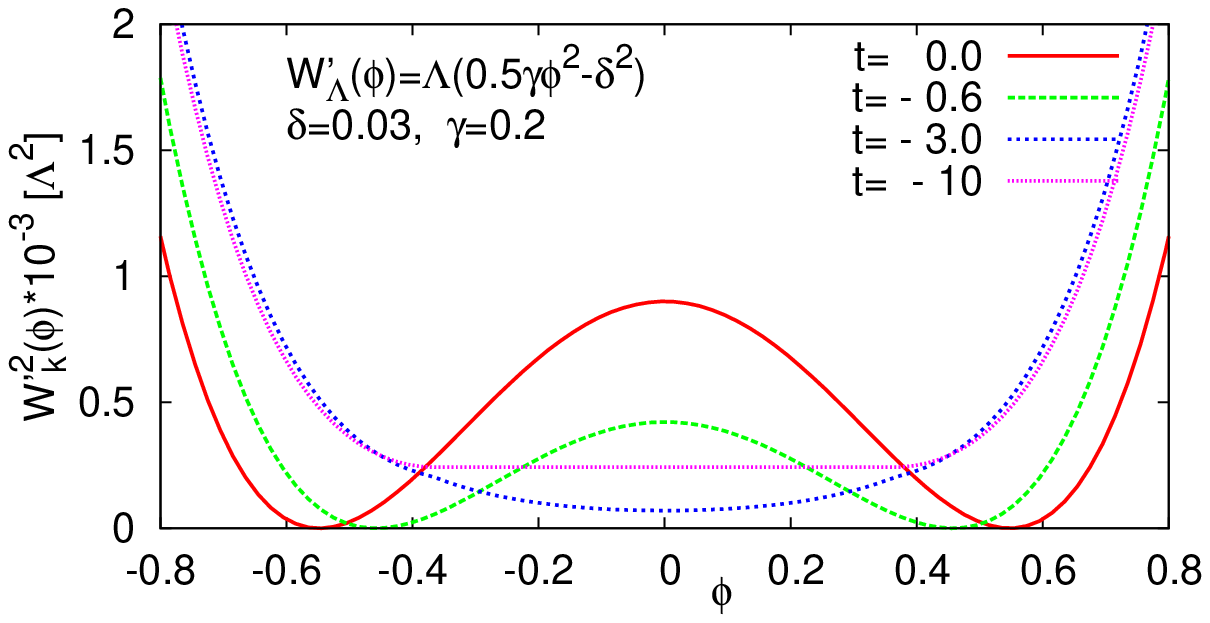}}
\vspace*{-2\baselineskip}
\vspace*{-2\baselineskip}
\centering{
\includegraphics[width=.99\columnwidth]{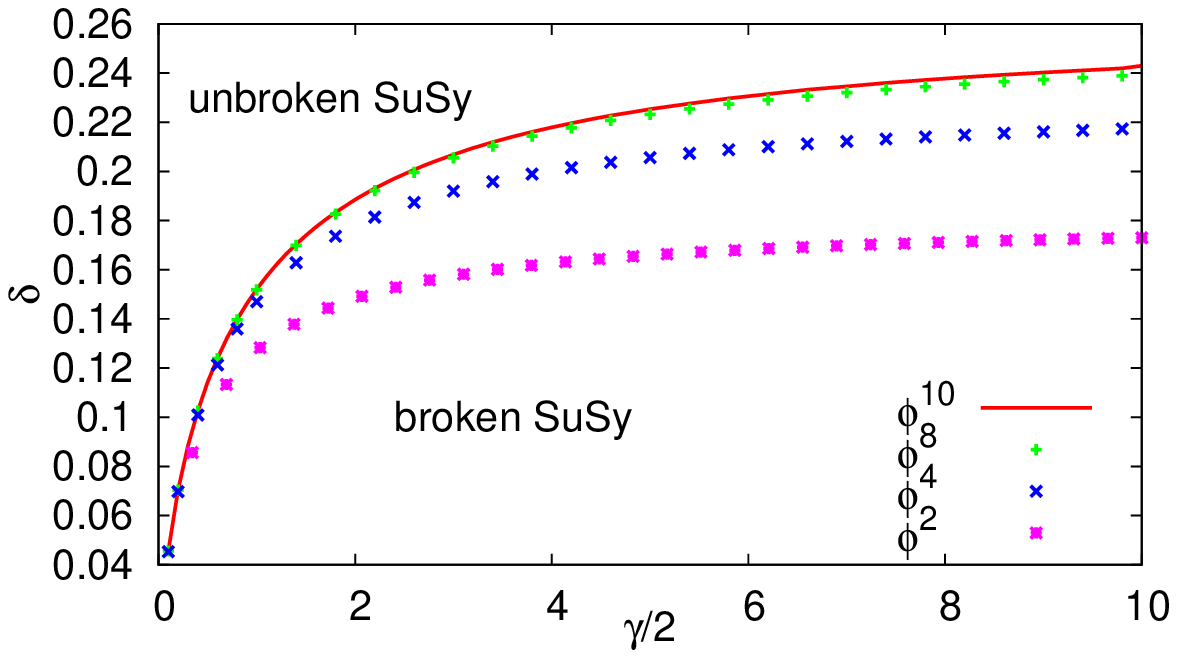}}
\vspace*{-2.5\baselineskip} 
\caption{\emph{Top:} Typical flow of the superpotential in the supersymmetry
broken phase. \emph{Bottom:} Phase diagram in the coupling--control-parameter
plane for the Gau\ss ian Wess-Zumino model. The lowest set of data points corresponds to the lowest-order result \eqref{eq:phase3}. Higher orders converge
  rapidly. \label{fig:Phas}}
\end{figure}

In order to clarify the role of this relevant direction and its relation to
supersymmetry breaking, we study the phase diagram of a Gau\ss ian Wess-Zumino
model at infinite volume. This is defined in terms of a quadratic perturbation
of the Gau\ss ian fixed point at the UV cutoff $\Lambda$:
$W_\Lambda'=\bar\lambda_\Lambda (\phi^2 - \bar a_\Lambda^2)$.  The
dimensionless control parameter $\delta$ and coupling $\gamma$ can be
associated with
\begin{equation}
\delta:= \Lambda^{-1}\, {\bar\lambda_\Lambda \bar a_\Lambda^2}, \quad \gamma :=
{2 {\Lambda}^{-1}\, \bar\lambda_\Lambda}, 
\end{equation}
where the scale is set by the UV cutoff $\Lambda$.  In the following, we
determine the region for the parameters $\delta,\gamma$ for which
supersymmetry is dynamically broken in the IR. A criterion for supersymmetry
breaking is provided by a nonvanishing ground state energy, given by the
minimal value of $V(\phi)=\frac12(W_{k\to 0}'(\phi))^2$. As a true phase
transition only occurs in the infinite-volume limit \cite{Witten:1982df}, the ground state
energy can only be identified at $k\to 0$. 
Four snapshots of a typical flow of the superpotential in the broken phase are
depicted in Fig.~\ref{fig:Phas} (top) for $\delta=0.03$ and $\gamma=0.2$.
      
For a simple estimate of the phase diagram, we truncate the
superpotential at order $\phi^2$ in \Eqref{eq:poly}. From \Eqref{eq:flow3}, we
obtain the flow of the dimensionless couplings
\begin{align}
		\partial_t a^2
		&=\frac{1}{2\pi }-\frac{6\lambda^2 a^2}{\pi },\quad
	\partial_t\lambda=-\lambda+\frac{6\lambda^3}{\pi},
	\label{eq:phase1}
\end{align}
which can be solved analytically. The phase transition curve in the
$(\gamma,\delta)$ plane is given by
the initial values for which $ a^2|_{k\to 0}$ changes sign.  This condition
yields
\begin{align}
\delta=\frac{\arcsin (\al)}
{\sqrt{24\pi}\,\al},\qquad
\al^2=1-\frac{\pi}{3\gamma}.
\label{eq:phase3} 
\end{align}
Beyond this simple estimate, the phase transition curve can, of course, be
identified by a full numerical integration of the flow equation. Already upon
inclusion of higher orders of \Eqref{eq:poly}, we observe a rapid convergence
of the phase boundary; see Fig.~\ref{fig:Phas} (bottom).  There exists a
critical value for $\delta_{\text{cr}}$ characterizing the phase transition in
the strong coupling limit $\gamma\to\infty$.  Our best estimate from a
numerical higher-order solution is $\delta_{\text{cr}} \simeq0.263$; (the
lowest-order result from Eq.~\eqref{eq:phase3} is $\sqrt{\pi/96}\simeq
0.181$).  We conclude that supersymmetry can never be broken dynamically above
this critical value. This agrees qualitatively with earlier results in the
literature \cite{Witten:1982df,Ranft1984166,Beccaria:2004ds,Beccaria:2004pa}.
Of course, the numerical value for $\delta_{\text{cr}}$ is not universal but
regulator-scheme dependent, simply because the quadratic initial conditions
near the Gau\ss ian fixed point are nonuniversal. A quantitative comparison
with other estimates of this critical value, e.g., taken from the lattice, is
thus only meaningful if this scheme dependence is accounted for. 

The mass of the lowest bosonic excitation is given by the curvature of the
full renormalized effective potential at the minimum. In the supersymmetric
phase, fermionic and bosonic masses are identical and nonzero. In
the broken phase, the lowest fermionic excitation is a Goldstino with
vanishing mass. The latter holds also at finite $k$ in the
supersymmetry-broken regime. As the flow in the IR is attracted by the
maximally IR-stable fixed point, the bosonic mass in the fixed-point regime is
governed by the only relevant direction $a^2$. As the minimum of the effective
potential in the broken regime is necessarily at a vanishing field, the bosonic
mass yields
\begin{equation}
m_k^2= 2k^2 \lambda^2 |a^2| \sim k^{2-\frac{1}{\nu_W}}, \label{eq:mass}
\end{equation}
where the latter proportionality holds in the vicinity of the IR fixed point
where $\lambda\to$const. and $a^2$ is governed by the critical exponent
$\nu_W$.  For $\nu_W>1/2$, the bosonic mass scales to zero upon attraction by
the maximally IR-stable fixed point. We conclude that the broken phase remains
massless in both bosonic and fermionic degrees of freedom.

\begin{figure}
\vspace*{-1.5\baselineskip} 
\centering{ 
\includegraphics[width=.99\columnwidth]{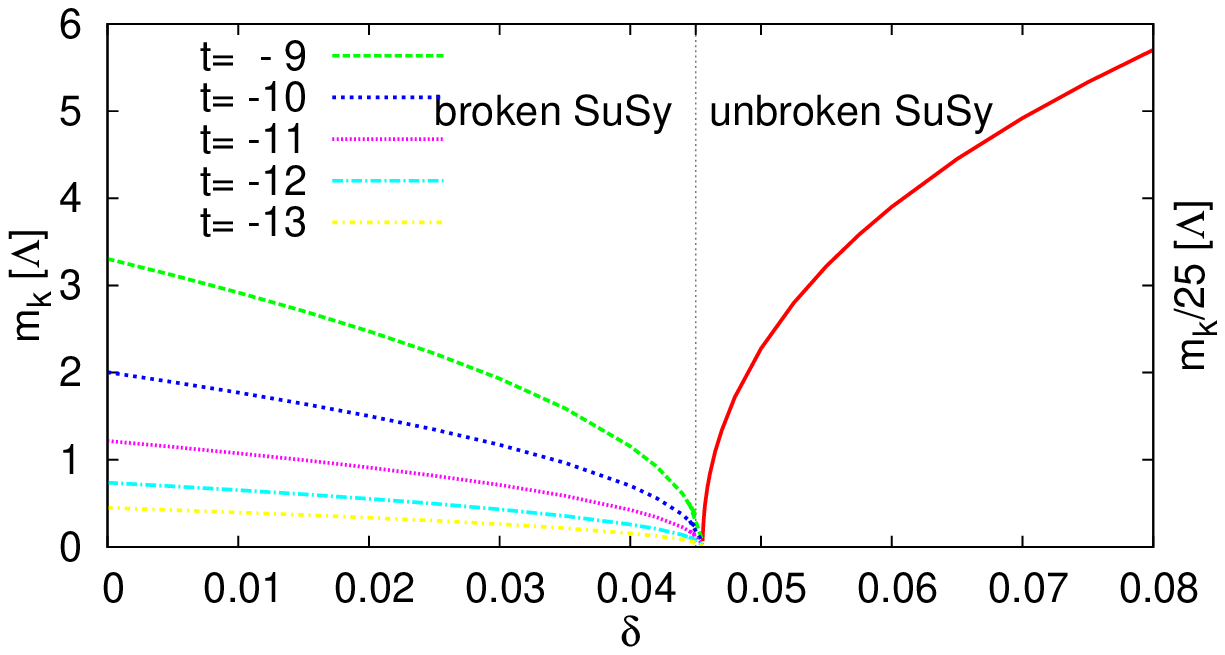}}
\vspace*{-2\baselineskip}
\vspace*{-2\baselineskip}
\centering{
	\includegraphics[width=.99\columnwidth]{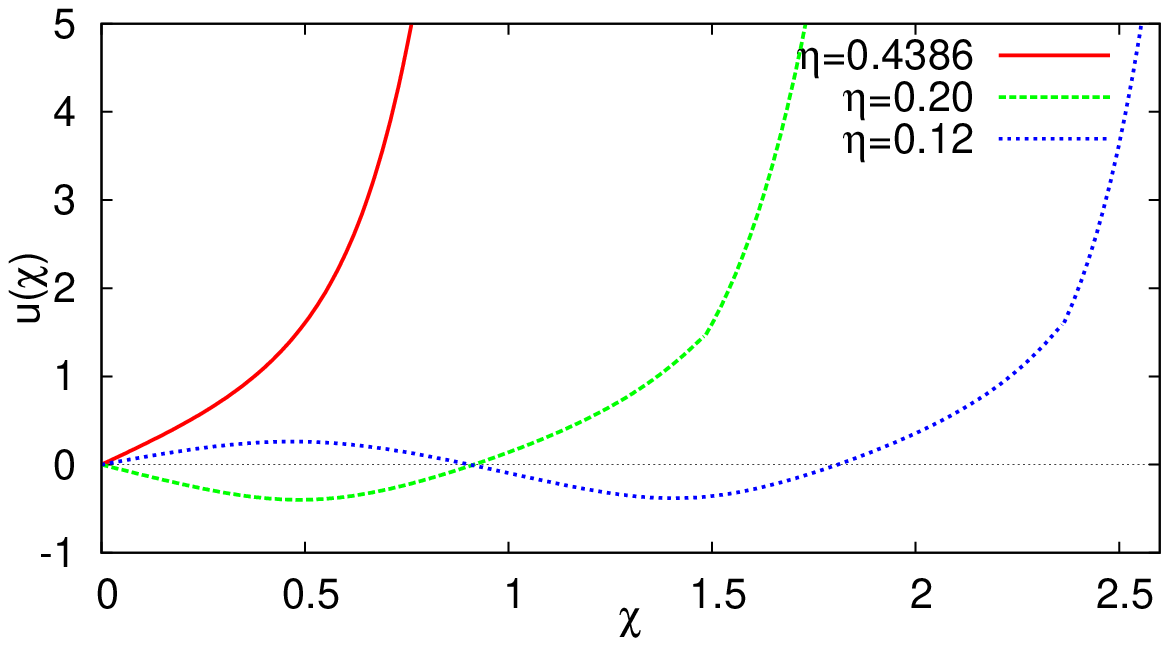}}
\vspace*{-2.5\baselineskip}
	\caption{\emph{Top:} Scale dependence of the renormalized bosonic mass.
	\emph{Bottom:} Fixed-point superpotentials at next-to-leading order derivative
	expansion, $\chi=Z_k \phi$, $u=w_k''/Z_k^2$.}
	\label{fig:RegPotential} 
\end{figure}

On the other hand, the limit $k\to0$ is an idealized IR limit. Any real
experiment as well as any lattice simulation will involve an IR cutoff scale
$k_{\text{m}}$ characterizing the measurement, e.g., the scale of momentum
transfer, detector size or lattice volume. Any measurement is therefore not
sensitive to $k\to 0$ but to $k\to k_{\text{m}}>0$. We conclude that a
measurement of the bosonic mass in the broken phase will give a nonzero answer
proportional to the measurement scale, whereas the goldstino will be truly
massless. The bosonic mass in both phases as a function of the control
parameter $\delta$ is depicted in Fig.~\ref{fig:RegPotential} (top) for a
coupling of $\gamma=0.2$.

At next-to-leading order in the derivative expansion,
the fields acquire a wave function renormalization $Z_k$, which is a
multiplicative factor of the kinetic term in \Eqref{eq:kin},
$\mathcal{L}_{\text{kin}} \to Z_k^2 \mathcal{L}_{\text{kin}}$.  The flow of
$Z_k$ gives rise to an anomalous dimension, $\eta = -\partial_t \ln
Z_k^2$. Convergence of the derivative expansion requires $\eta\lesssim
\mathcal O(1)$. 

From the polynomially expanded superpotential flow at next-to-leading order
we can derive the following flow of the renormalized parameter $a^2$:
\begin{equation}
\partial_t a_t^2 = \frac{1}{2\pi} \left( 1- \frac{\eta}{4} \right) - \left( 1-
\frac{\eta}{2} \right) a_t^2 - \frac{a_t^2}{\lambda_t}\, \partial_t \lambda_t.
\end{equation}
At any fixed point with $\partial_t\lambda\to0$, we can read off that
$a^2$ again denotes a relevant direction with scaling $a^2 \sim k^{-1/\nu_W}$,
where 
\begin{align}
 	\nu_W=\frac{2}{2-\eta}.
 	\label{eq:scale1}
\end{align}
This remarkable {\em superscaling} relation connects the superpotential
exponent $\nu_W$ with the anomalous dimension $\eta$. Recall that in
Ising-like systems the thermodynamic main exponents (i.e., $\alpha$, $\beta$,
$\gamma$ and $\delta$) are related among each other by scaling relations, and
can be deduced from the correlation exponents $\nu$ and $\eta$ by hyperscaling
relations. Beyond that, there is no general relation between $\nu$ and
$\eta$. The superscaling relation \eqref{eq:scale1} thus represents a special
feature of the present supersymmetric model and is an exact relation to
next-to-leading order in the supercovariant derivative expansion.  Beyond
next-to-leading order, Eq.~\eqref{eq:scale1} might receive corrections, since
higher-derivative operators involving the auxiliary field can still take
influence on the flow of the superpotential. Nevertheless, even if
\Eqref{eq:scale1} receives higher-order corrections, the most important result
that a potentially more involved scaling relation among $\nu_W$ and $\eta$
exists still holds. 

\begin{table}
 \begin{ruledtabular} 
\begin{tabular}{ccc}
\# nodes&$\eta$&${\nu_W}$\\\hline
0&0.4386&1.2809\\
1&0.20	&1.11\\
2&0.12	&1.06\\
\end{tabular}
\end{ruledtabular}
\caption{Critical exponents of the first fixed points.\label{tab:exponents}}
\end{table} 

Finally, we determine the possible nontrivial fixed-point
superpotentials. Solving the coupled fixed-point equations for the
dimensionless superpotential as well as the anomalous dimension
self-consistently, we find a discrete set of fixed-point superpotentials with
an increasing number of nodes, see Fig.~\ref{fig:RegPotential} (bottom). The
number of RG relevant directions at the fixed points in addition to the $a^2$
direction is equal to the number of nodes. The superpotential with no nodes is
the next-to-leading order analogue of the maximally IR-stable fixed point
discussed above. Our quantitative next-to-leading order estimates of the
critical exponent $\nu_W$ and the anomalous dimension $\eta$ for the
fixed-point superpotentials of Fig.~\ref{fig:RegPotential} (bottom) are
summarized in Tab.~\ref{tab:exponents}. As these quantities are universal,
these results should constitute regulator-independent predictions. Of course,
the truncation of the effective action introduces an artificial regulator
dependence, the size of which can be used for an error estimate of our
numerical results. From our regulator studies, we found a maximum variation of
10\%\ on the positive critical exponents. Subleading negative exponents can
show larger variations and thus are difficult to estimate reliably, whereas
the leading exponents vary even less, if at all \cite{Synatschke:2009}. 

Each fixed point constitutes a universality class and defines a distinct
asymptotically safe UV completion of the Wess-Zumino model, i.e., RG
trajectories emanating from different fixed points correspond to different
theories. Our findings are reminiscent of those in 2D scalar field theories
\cite{Morris:1994jc,Neves:1998tg}, where the fixed points of the effective
potential can be related to conformal field theories \cite{Zamolodchikov86}.
The connection between the present supersymmetric models at their fixed points
and conformal field theories remains an interesting question.

To summarize, we have constructed a manifestly supersymmetric flow equation
for 2D $\mathcal N=1$ Wess-Zumino models.  These models have
nontrivial fixed-point superpotentials which can be classified by their
relevant directions. All fixed points share a relevant
direction in the form of a bosonic mass term.  At next-to-leading order, the
associated critical exponent $\nu_W$ and the anomalous dimension $\eta$
satisfy a superscaling relation which has no analogue in Ising-type spin
systems. 

The fixed point of this relevant direction corresponds to the critical point
separating the supersymmetric from the broken phase. The initial condition for
this relevant direction defines a control parameter, such that supersymmetry
breaking can be understood as a quantum phase transition.  For the Gau\ss ian
Wess-Zumino model, we observe that the control parameter stays finite even at
arbitrarily large coupling in accord with a general argument by Witten. In the
broken phase, our superscaling relation predicts that the measured bosonic
mass is proportional to the momentum scale set by the detector. 

\acknowledgments{Helpful discussions with G.~Bergner, C.~Wozar, T.~Fischbacher
  and T.~Kaestner are gratefully acknowledged.  This work has been supported
  by the Studienstiftung des deutschen Volkes, and the DFG under GRK 1523,
  Wi 777/10-1, FOR 723 and Gi 328/5-1.}

\end{document}